\documentclass{article}

\usepackage{arxiv}

\usepackage[utf8]{inputenc} 
\usepackage[T1]{fontenc}    
\usepackage{hyperref}       
\usepackage{url}            
\usepackage{booktabs}       
\usepackage{amsfonts}       
\usepackage{nicefrac}       
\usepackage{microtype}      
\usepackage{lipsum}		
\usepackage{graphicx}
\usepackage{hyperref}
\hypersetup{colorlinks,allcolors=black}
\usepackage{doi}
\usepackage{graphicx}
\usepackage[justification=centering]{caption}
\makeatletter
\newcommand*{\rom}[1]{\expandafter\@slowromancap\romannumeral #1@}
\makeatother
\usepackage{lineno}
\usepackage{float}
\usepackage{amsmath, amsthm, amssymb}

\usepackage{lineno}
\usepackage{lineno,hyperref}
\usepackage{siunitx}
\usepackage{makecell}
\usepackage{pbox}
\usepackage{fourier} 
\usepackage{array}
\usepackage{float}
\usepackage{lscape}
\usepackage[table,xcdraw]{xcolor}
\usepackage{breqn}

\usepackage{amsmath}
\usepackage{amsmath,amssymb}
\usepackage{textcomp}
\usepackage{graphicx}
\usepackage[justification=centering]{caption}
\usepackage{lineno}
\usepackage{float}
\usepackage{amsmath,amssymb}
\usepackage{amsthm}
\usepackage{lineno,hyperref}
\usepackage{siunitx}
\usepackage{makecell}
\usepackage{pbox}
\usepackage{fourier} 
\usepackage{array}
\usepackage{xcolor}
\usepackage{float}
\usepackage{lscape}
\SetMathAlphabet{\mathtt}{normal}{OT1}{SourceCodePro-TLF}{m}{n}

\usepackage[table,xcdraw]{xcolor}

\title{A Bayesian Neural Network Approach for Tropospheric Temperature Retrievals from a Lidar Instrument}

\author{ \href{}{\hspace{1mm}Ghazal Farhani}\\
	National Research Council Canada\\
	Automotive and Surface Transportation Research Centre\\
	London, ON, Canada \\
	\texttt{Ghazal.Farhani@nrc-cnrc.gc.ca} \\
	\And
\href{}{Giovanni Martucci} \\
	Federal Office of Meteorology and Climatology\\
	MeteoSwiss\\
	Payerne, Switzerland \\
	\texttt{Giovanni.Martucci@meteoswiss.ch} \\
	\And
\href{}{ Tyler Roberts} \\
	Department of Physics and Astronomy\\
	The University of Western Ontario\\
	London, ON, Canada  \\
	\texttt{tylergordonr@gmail.com } \\
	\And
\href{}{ Alexander Haefele} \\
	Federal Office of Meteorology and Climatology\\
	MeteoSwiss\\
	Payerne, Switzerland \\
	\texttt{Alexander.Haefele@meteoswiss.ch} \\
	\And
\href{}{Robert J. Sica} \\
	Department of Physics and Astronomy\\
	The University of Western Ontario\\
	London, ON, Canada  \\
	\texttt{sica@uwo.ca } \\
 }

\renewcommand{\shorttitle}{A Bayesian Neural Network Approach for Tropospheric Temperature Retrievals from a Lidar Instrument}

\hypersetup{
pdftitle={A template for the arxiv style},
pdfsubject={q-bio.NC, q-bio.QM},
pdfauthor={Ghazal Farhani, Giovanni Martucci, Tyler Roberts, Alexander Haefele and Robert J. Sica},
pdfkeywords={First keyword, Second keyword, More},
}

\begin{document}
\maketitle

\begin{abstract}
We have constructed a Bayesian neural network able of retrieving tropospheric temperature profiles from rotational Raman-scatter measurements of nitrogen and oxygen and applied it to measurements taken by the RAman Lidar for Meteorological Observations (RALMO) in Payerne, Switzerland. We give a detailed description of using a Bayesian method to retrieve temperature profiles including estimates of the uncertainty due to the network weights and the statistical uncertainty of the measurements. We trained our model using lidar measurements under different atmospheric conditions, and we tested our model using measurements not used for training the network. The computed temperature profiles extend over the altitude range of 0.7\,km to 6\,km. The mean bias estimate of our temperatures relative to the MeteoSwiss standard processing algorithm does not exceed 0.05\,K at altitudes below 4.5\,km, and does not exceed 0.08\,K in an altitude range of 4.5\,km to 6\,km. This agreement shows that the neural network estimated temperature profiles are in excellent agreement with the standard algorithm. The method is robust and is able to estimate the temperature profiles with high accuracy for both clear and cloudy conditions. Moreover, the trained model can provide the statistical and model uncertainties of the estimated temperature profiles. Thus, the present study is a proof of concept that the trained NNs are able to generate temperature profiles along with a full-budget uncertainty. We present case studies showcasing the Bayesian neural network estimations for day and night measurements, as well as in clear and cloudy conditions. We have concluded that the proposed Bayesian neural network is an appropriate method for the statistical retrieval of temperature profiles.
\end{abstract}

\keywords{ Atmospheric Temperature; Neural Networks; Bayesian Deep Learning; Raman Lidar; Atmospheric Retrievals }

\section{Introduction}
Enhancing our knowledge of trends and variability in atmospheric temperature is essential to better understanding weather and climate change's impacts and causes\cite{santer1996search, tett1996human, thorne2011tropospheric, philipona2018radiosondes}. Tropospheric warming has been measured by different observational platforms since the mid-twentieth century. However, the confidence level in the rate of its change and its vertical structure remains relatively low, which can limit the ability to produce reliable inferences about the true long-term trends. Continuous measurements along with developing reliable methods of retrieving temperature profiles are vital for climate change studies. Pure Rotational Raman (PRR) lidars are ground-based remote sensing instruments with excellent vertical and temporal resolutions, suitable to provide accurate tropospheric temperature profiles \cite{cooney1972measurement,behrendt2005temperature, mahagammulla2019retrieval}. Typically, to retrieve temperature profiles the ratio of measurements from two pre-processed PRR signals is calculated. The PRR spectrum contains two symmetrically positioned Stokes and anti-Stokes branches on either side of the excitation line with approximately the same intensity \cite{cooney1972measurement}. The pre-processing of PRR signals involves removing the background counts and implementing saturation corrections. Depending on the system it might be essential to consider some additional height-dependent corrections. Moreover, external measurement of temperature (typically from a coincident radiosonde) is required to find the coefficients of the calibration function, which relates the lidar rotational-Raman-intensity measurements to temperature profiles \cite{behrendt2005temperature}. Here we demonstrate that neural network (NN) algorithms are able to accurately retrieve Raman temperature profiles. Moreover, we present a simple yet effective method to calculate the statistical and model uncertainties of the estimated temperature profiles. 

Machine learning methods, specifically, NNs have recently become popular among researchers in different fields from network security to medicine. Some interesting implementations of this technique can be found in \cite{pierson2017deep, duc20203d, yazdi2020sefee, sun2020surrogate}. NNs have also been implemented for atmospheric constituent retrievals using different instruments \cite{muller2003ozone, blackwell2005neural, del2005neural, milstein2016neural, cai2020temperature}. These algorithms are capable of learning complex relations between input and output data without a need to know the mapping function. Moreover, to train a NN algorithm there is no need to perform any data pre-processing steps needed for the traditional analysis. Although, NNs are powerful predictive algorithms, often they do not quantify the uncertainty of the output. For atmospheric temperature profiles, knowing the level of uncertainty of the output is essential. The uncertainty of the model parameters (weights) needs to be calculated. Quantifying uncertainty in NNs is an area of ongoing research, and many studies have been conducted to address the issue \cite{tiwari2010uncertainty, blundell2015weight, gal2016dropout, lakshminarayanan2016simple, pearce2020uncertainty}. 

Two major contributions of this study are to explore the possibility of implementing NNs to estimate the temperature profiles, and to quantify the uncertainties of the estimated temperature profiles. By implementing NNs, we can avoid the data pre-processing tasks which are typically needed in conventional temperature retrievals. Although none of the pre-processing tasks are difficult, they can bear large uncertainties, thus all the independent uncertainty components should be calculated and correctly propagated through the data processing chain \cite{leblanc2016proposed}. For example, in many lidar systems, the raw profiles from different measurement channels are glued, the merging process is empirical and can be a source of uncertainty hard to be calculated \cite{zhang2014slope}. Also, it is possible to encounter the signal-induced noise (SIN) caused by high photon counts. SIN should be modeled correctly to remove background counts. The Modeling of SIN is yet based on empirical methods and bears uncertainties \cite{pettifer1975signal, acharya2004signal}. Furthermore, estimating some characteristics of the lidar instruments such as the lidar overlap function can be challenging, and mostly is based on empirical methods \cite{wandinger2002experimental, hervo2016empirical}. Thus, NNs have the potential of providing temperature profiles without the need for data pre-processing steps. Moreover, although NNs have been widely used for retrieving atmospheric constituents, to our knowledge, this is the first attempt to include model uncertainties for each retrieved profile, and a complete profile-based uncertainty budget is provided. Here we use RALMO measurements as input and the traditional retrievals from RALMO as the ground truth to train a Bayesian NN algorithm to retrieve tropospheric temperature profiles and to provide its uncertainty. 
In Sect. 2, we present a brief description of the instrument. We also discuss the traditional method of calculating temperature profiles. Sect. 3, is a brief description of NN models and methods of quantifying their uncertainties.  In Sect. 4, we describe the general architecture of the NN which was built to train the temperature profiles. Sect. 5 includes full descriptions of three case studies and discusses the overall performance of the Bayesian NN algorithm for the entire test data-set. Sect. 6 is a summary of the results, and short discussion on evaluating the implementation of the Bayesian NN and its advantages and shortcoming for retrieving temperature profiles. It also provides a future map toward using NN algorithms in atmospheric studies. 


\section{RALMO}

\subsection{RALMO}
The RAman Lidar for Meteorological Observations (RALMO) is located in Payerne Switzerland ($46.81^{\circ}$N, $6.94^{\circ}$E, 491\,m ASL). The lidar was built at the École Polytechnique Fédérale de Lausanne and is being operated by the Federal Office of Meteorology and Climatology, hereafter referred to as MeteoSwiss \cite{dinoev2013raman}. RALMO has been fully operational since 2008 and has provided nearly continues measurements of the tropospheric temperature since then. In clear atmospheric conditions,  with 30 minute integration time, it is capable of reaching to 6-7\,km during day and 12\,km during night time measurements. The lidar is equipped with a frequency-tripled Nd:YAG laser emitting at 355\,nm with an emission energy of about 400\,mJ per pulse and the repetition rate of 30\,Hz. Using a beam expander, the beam's diameter is expanded to 14\,cm which reduces the beam divergence to 0.09 $\pm$ 0.02\,mrad. In the receiving end of RALMO, four high-efficiency reflecting parabolic mirrors with 30\,m diameters are used to collect the backscattered signals. The Raman-shifted backscattered signals are received after passing through a two-stage polychromator diffracting process, and then recombined into two groups of $J_{high}$ and $J_{low}$ signals. The $J_{high}$ and $J_{low}$ signals represent the high and low quantum number lines in the Stokes and anti-Stokes branches respectively. Details on RALMO instrumentation, updates and its characteristics can be found in \cite{martucci2021validation}. 

\subsection{Traditional Temperature Retrievals}

The interaction between the emitted light from laser and ${O_2}$ and ${N_2}$ molecules in the atmosphere results in a frequency-shifted Raman signal which is back-scattered to the lidars's receiver. The measured backscattered signal at altitude $z$ over time $t$ is given by the Raman lidar equation:     
\begin{equation}  \label{lidar equation}
    S(z) = \frac{C}{z^2} O(z) n(z) \mathcal{L}_{atm} ^2 (z) \left[\sum_{i = O_2, N_2} \sum_{J_i} \tau(J_i) \eta_i (\frac{d\sigma}{d\Omega})^i (J_i)\right] + B
\end{equation}
where $C$ is the lidar constant, $O(z)$ is the geometrical overlap between transmitted laser beam and telescope, $\textit{n(z)}$ is the air number density, $\mathcal{L}_{atm}^2(z)$ is the atmospheric round-trip transmission, $\tau(J_i)$ is the transmission at the wavelength of each line $J_i$, $\eta_i$ is the volume mixing ratio for each molecule, $(\frac{d\sigma}{d\Omega})^i (J_i)$ is the Raman cross section for each $J_i$, and $B$ is the background counts. The high frequency-shifted and low frequency-shifted signals are described by the lidar equation, where their ratio $Q(z)= \frac{J_{low}(z)}{J_{high}(z)}$ is equivalent to:
\begin{equation}\label{Q_ratio}
    Q(z) = \frac{\left[\sum_{i = O_2, N_2} \sum_{J_i} \tau(J_i)_{low} \eta_i (\frac{d\sigma}{d\Omega})^i (J_i)\right]}{\left[\sum_{i = O_2, N_2} \sum_{J_i} \tau(J_i)_{high} \eta_i (\frac{d\sigma}{d\Omega})^i (J_i)\right]}.
\end{equation}
In theory, by calculating the differential backscatter cross-section area for the two wavelengths, and inserting the transmission values at each ${J_i}$, the temperature profile can be calculated. However, in practice, temperature profiles retrieved from a nearby radiosonde is used, and $Q$ is related to the temperature profile as: 
\begin{equation} \label{Q}
    T = \frac{A}{B + \ln{Q}}
\end{equation}
where $\textit{T}$ is the temperature profile and $\textit{A}$ and $\textit{B}$ are two coefficients which are determined by calibrating $\textit{T}$ with respect to coincident radiosonde temperature measurements. It is important to note that in order to retrieve $\textit{T}$ correctly, before calculating $\textit{Q}$, the signals should be corrected for dead time of the acquisition system and the background counts. A detailed description on the steps towards calculating the temperature profiles is available in \cite{behrendt2005temperature}. 

Recently, \cite{martucci2021validation} showed that the temperature profiles retrieved from RALMO data during the period of July 2017 to December 2018 were in an excellent agreement with the two daily co-located reference, radiosonde flights. The mean difference between the traditional and the radiosonde profiles in daytime was found to be $0.62 \pm 0.1$\,K and in nighttime was reported to be 
$0.66 \pm 0.34$\,K. Their result indicated that the RALMO temperature profiles retrieved using this procedure had high stability and small statistical uncertainty. Thus, we have used the traditional temperature profiles and their corresponding statistical uncertainty profiles as the ground truth profiles in the study. We have used the same period of lidar measurements and their corresponding retrieved temperatures to train our NN algorithm. 


\section{Neural Networks for Regression}

\subsection{Neural Networks}
NNs are ensemble of nonlinear functions that are trained to infer a statistical relationship between the input and output vectors from a set of training examples. NNs have a layer-wise structure containing simple computational elements known as nodes. At each layer, nodes are connected to the previous and to the next layer. The connections are weighted. A weight matrix connects one layer to the next layer. Formally, output of layer $j$ can be written as: 
\begin {equation}
\psi_{j} (\sum_{i}^N w_{ji}x_{i} + b_{j}) =
\psi_{j} (\mathbf{W}^\top \mathbf{x}+ b)
\end{equation}
where N is the number of neurons in the layer, $\psi$ is a nonlinear function applied to the weighted sum of input vector (\textbf{x}) and a weight matrix (\textbf{W}). The nonlinear function which is known as activation function is chosen from Sigmoidal functions. Sigmoidal functions are real-valued functions such that:
\begin{equation}
\begin{split}
\lim_{x \to -\infty} \psi(x) =&0 \\
\lim_{x \to \infty} \psi(x) =&1.
\end{split}
\end{equation}

\begin{figure}[h!]
\centering
\includegraphics[width= 9cm, height = 6 cm]{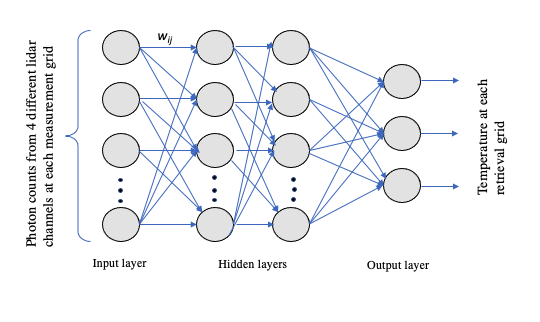}
\caption{The architecture of an NN with an input layer, 2 hidden layers, and an output layer. The layers are connected via elements of the weight matrix \textbf{W}. The input and the output do not need to be the same grid.}
\label{fig: NN_schematics}\end{figure}

The architecture of a NN with an input layer, 2 hidden layers, and an output layer is shown in Fig.~\ref{fig: NN_schematics}. In the context of lidar measurements, the input data is a vector containing photon counts at each altitude and for each channel. For temperature retrievals, an output layer is a vector containing the temperature at each altitude. Layers between the input and output layers are called hidden layers, and they represent the depth of NN.

During the training process, the weights and biases are tuned to minimize a cost function. For regression, normally mean squared error between the target vector \textbf{y} and the prediction $\hat {\textbf{y}}$ is chosen:

\begin {equation}
L_{\mathbf{w}}(\mathbf{w}, \mathbf{x}, \mathbf{y}) = \frac{1}{N} \sum_i (y_i - {\hat{y_i}})^2.
\end{equation}
The cost function is numerically minimized with respect to the parameters. Most optimization algorithms are similar to the gradient descent with the following update rule:

\begin {equation}
w_{i, t+1} = w_{i,t} - \tau  \frac{\partial L}{\partial w_{i,t}}
\end{equation}
where $t$ is the current time and $\tau$ is a learning rate that defines the size of update, and is set by the user \cite{zhigljavsky2007stochastic, torn1999stochastic}. In practice, in each time step, instead of computing the gradient of the whole data-set, the gradient of a subset of data is being calculated; the later method is known as stochastic gradient descent. Adam is another variant of the gradient descent in which a moving average of the gradient mean and variance is stored. This approach leads into more effective weight updates \cite{kingma2014adam}.

\subsection{Quantifying Uncertainty in Neural Networks}
As mentioned in Section 3.1, the optimized target function is the result of optimized weights. Thus, to calculate the effect of model uncertainty on target functions, the uncertainty of weights should be calculated. Quantifying the uncertainty of NNs' predictions for physical systems is essential. One major approach to adopt uncertainties for NN models is to use Bayesian formalism in which parameters (weights) of a NN are set to follow \textit{apriori} Normal distribution, then the \textit{posterior} distribution over the parameters is computed that yields in estimating model uncertainty. Using Bayes' theorem, given a training data-set, the \textit{posterior} distribution for the space of parameters is written as \cite{bishop1995training, theodoridis2015machine}:
\begin {equation} \label{Bays's theorem}
P(\mathbf{W} | {\mathbf{x}},{\mathbf{y}}) = \frac{P(\mathbf{W}) P({\mathbf{y}}|\mathbf{W},{\mathbf{x}})}{P({\mathbf{y}}|{\mathbf{x}})}
\end{equation}
where $P({\textbf{y}}|{\textbf{x}}) $ is calculated as:
\begin {equation} \label{mergenalize}
P({\mathbf{y}}|{\mathbf{x}}) = \int P({\mathbf{y}}|{\mathbf{x}} ,\mathbf{w}) P(\mathbf{w}) d \mathbf{w}
\end{equation}
Performing the integration in Eq.\ref{mergenalize} is called marginalising the likelihood over the parameter. Using Eq.\ref{Bays's theorem} we can predict an output for an unseen input data point ${\textbf{x*}}$:
\begin {equation} \label{Inference}
P({\mathbf{y*}}|{\mathbf{x*}}, {\mathbf{x}},{\mathbf{y}}) = \int P(w | {\mathbf{x}},{\mathbf{y}}) P({\mathbf{y*}}|{\mathbf{x*}},{w})dw
\end{equation}
The integral of Eq. \ref{Inference} is called Inference; for most models, the marginalising cannot be done analytically, and an approximation is needed. A verify of methods have been developed to approximate the Bayesian inference among which Markov chains Monte Carlo (MCMC) methods \cite{quinonero2005evaluating} and its variants such as Langevin diffusion methods and Hamiltonian methods are well-known \cite{welling2011bayesian, korattikara2015bayesian, hmcspringenberg2016bayesian}. Recently, variational Bayesian methods have gain interest as well \cite{blundell2015weight}. Although, these approximations are appealing, computationally they are very expensive. 

Alternative to the Bayesian approach is ensemble methods in which the estimate of multiple individual models will be aggregated. Randomization-based and boosting-based approaches are the main branches of ensemble methods. In randomization-based models each ensemble can be trained individually and independent of other ensembles. The variance of the ensembles' predictions is interpreted as the uncertainty of the prediction. Recently, \cite{lakshminarayanan2016simple} used the idea of ensemble for NNs. In their approach, they trained multiple individual NNs such that for each NN parameters are initialized randomly and data points are shuffled and randomly sampled. The model uncertainty was obtained by averaging predictions over multiple NN models. 

Recently, \cite{pearce2020uncertainty} used Randomized maximum a \textit{posterior} sampling (RMS) method that combines the Bayesian and ensemble methods. In the mentioned approach, by adding a regularization term to the cost function a maximum a \textit{posteriori} (MAP) of the parameters is estimated that is the point estimate of Bayesian posterior. Injecting noise to either terms of the cost function and sampling repeatedly will produce a distribution of solutions which can be shown is a good estimation of the true \textit{posterior} distribution \cite{bardsley2014randomize}. In practice, the mean of the distribution is used as the optimum parameter state and its standard deviation indicates the uncertainty of the estimation. 

More formally, given a prior Normal distribution for the parameters $ P(\textbf{W}) = N(\mu_{prior}, \sum_{prior})$ the MAP can be estimated as:
\begin{equation}
    \mathbf{W}_{{MAP}} = arg \max_{x}{\mathbf{W}} (\log P({\mathbf{x}},{\mathbf{y}}|\mathbf{W}))- \frac{1}{2} || {\sum_{prior}}^{-\frac{1}{2}} (\mathbf{W} - \mu_{prior})||^2.
\end{equation}
In the absence of $\mu_{prior}$, the equation becomes the standard L2 regularization. In the RMS method 
$\mu_{prior}$ is a random draw from $N(\mu_{prior}, \sum_{prior})$. Thus the cost function for regression for each net can be written as:
\begin{equation}
    Cost_{j} = \frac{1}{N} || \frac{(\mathbf{y} -  \hat{{\mathbf{y}}})^2}{\sigma_j^2} ||  + \frac{1}{N} || \mathcal{H}^{\frac{1}{2}}(\mathbf{w}_{j} - \mathbf{w}_{RMS,j})|| ^2
\end{equation}
where the uncertainty of the ground truth is $\sigma_j^2$, and the diagonal of $\mathcal{H}$ is defined as $\frac{1}{\sigma_{prior}^2}$ and $j$ indicates number assigned to each net.

We modified the algorithms to accommodate the effect of input (raw photon counts) uncertainty on the weight optimization procedure. At relatively modest count rates (e.g. $\sim$20 or less), the distribution of uncertainties for lidar measurements tends to be Normal. Thus, it is possible to add small Gaussian perturbations into the measurements in each iteration. During the training, we sample repeatedly from both input data (photon counts) and the weight spaces. In this approach, we can capture the effect of noise on the estimation of weight parameters. We also repeated the process by implementing a more accurate assumption that photon counts follow the Poisson distribution \cite{liu2006estimating}, the difference between the results was not significant.
Adding Gaussian noise into the input data also has the benefit of acting as a regularization term (Tikhonov regulizer), helping to avoid overfitting \cite{bishop1995training}. Moreover, similar to \cite{gal2016dropout} approach at test time, a small noise was added to the measurements, for multiple times, and for each ensemble, then the temperature estimation was calculated. For each measurement, the averaged profile was used as the final estimation and its standard estimation was used as the uncertainty of the estimation. 

We will show that the precision of our retrievals is limited by the precision of our ground truth. In the case of having an optimal model with the minimal model uncertainty, the estimation of the temperature profile will be similar to the ground truth; however, the uncertainty of the ground truth should be added to the total uncertainty of the estimations. In the traditional method, the statistical uncertainty of each profile is calculated \cite{martucci2021validation}. Thus, we are able to simultaneously estimate both temperature profiles and their corresponding uncertainty and produce two outputs: the estimated profile and the estimated uncertainty of the ground truth. The final variance estimation can be written as follows \cite{sengupta2020ensembling}:
\begin{equation}
    \frac{1}{m} \sum_{j=1} \hat{\sigma_j}^2 + \frac{1}{m} \sum_{j=1} \hat{{\mathbf{y_j}}}^2 - (\frac{1}{m} \sum_{j=1} \hat{{\mathbf{y_j}}})^2,
\end{equation}
where $m$ is the number of ensembles and $\hat \sigma_j^2$ is the estimated variance of the ground truth uncertainty for each profile. 

Of critical importance is that, the estimated uncertainty of the ground truth is inherited from data and cannot be minimized. However, the model uncertainty can be minimized as more data is added to the training set. 



\section{Description of the data and model}
A total of 4510 temperature profiles from RALMO measurements between July 2017 and August 2018 are used in this study. We divided our data into training (70 \% of total), validation (10\% of total) and test sets (20\% of total). Of note, during the mentioned period only 8167 measurements were recorded. That is because for many hours (or even days) measurements were halted due to weather conditions.  Among the available measurements, only raw data corresponding to temperature profiles with valid values in the height range from 0.7\,km to 6\,km was selected. This corresponds to 4510 measurement and temperature profiles. Thus, many temperature profiles that did not reach the pre-defined height were discarded. In the mentioned period (July 2017 to August 2018), if the lidar was fully operational, we would have more than 15000 raw profiles. Considering the weather conditions, as well as our criteria of selecting profiles, we had 4510 profiles (30\% of possible measurements). Thus, the dataset was already extremely sparse in time. However, to ensure that the training and test data were not correlated in time, we made sure that the test data were not selected from the same days (and nights) that we selected the training set.

The dimensions of the input layer corresponding to photon counts from $J_{high}$ and $J_{low}$ channels was 4688, and the output layer corresponding to the temperature profiles had 175 neurons. To improve the optimization process in NN models, it is often necessary to ensure that input data have similar ranges of values, thus the data should be scaled. Hence, each input measurement profile is standardized using the following equation:
\begin{equation} \label{data_standard}
    x_{standard} = \frac{x-\mu}{\sigma},
\end{equation}
where $\mu$ is the mean and $\sigma$ is the standard deviation of a profile. As mentioned earlier, we used temperature profiles from the traditional retrievals as the ground truth. During the period considered the lidar was calibrated 7 times, with only small differences between the calibration values. For more extended periods of measurements, or in conditions when the calibration constants vary drastically, it is possible to use $Q$ values as the ground truth and train the model to estimate the $Q$ values. Then, using Eq. \ref{Q} the temperature profile can be calculated. 

To build a RMS algorithm, we modified source codes developed by \cite{pearce2020uncertainty}. Our RMS model is trained using the Keras API in Python. The NN models have several hyperparameters which are not determined from the data during the training. The hyperparameter tuning is performed separately. The number of hidden layers, number of neurons per layer and choice of activation functions are important hyperparameters for a network. We trained our model multiple times using a set of random selections from different choices of activation functions, as well as different numbers of hidden layers and neurons. Considering both performance and the time needed to train the model, we constructed our final network using 4 hidden layers with each layer containing 250 neurons. A ``$\tanh$'' activation function is used to connect the hidden layers. Using denser NNs was computationally expensive, and did not improve the estimations.  Estimating the error for all training samples leads to a slow convergence. Thus we used a mini-batch approach at each epoch. The batch size was another hyperparameter, and batch size of 128 was selected. As the mini-batch approach can lead to more noisy estimates, we used an Adam algorithm, which is a well suited optimization algorithm for noisy data \cite{kingma2014adam}. We also used a stochastic gradient algorithm which had a lower performance compared to Adam.

\section{Results}

\begin{figure}[htb!]
\centering
\includegraphics[scale=.32]{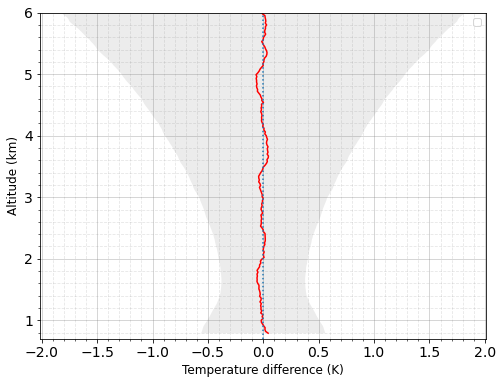}
\includegraphics[scale=.46]{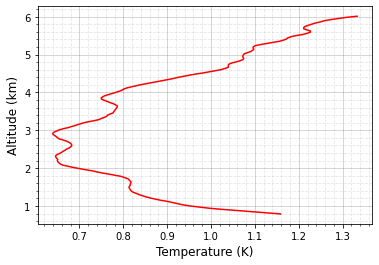}
\caption{Left panel: The average of MBE (red line) is shown. The shaded gray area represents the average of the summation statistical uncertainty of the target profiles. Right panel: RMSE calculation for the temperature profiles on the test data set. }
\label{MBE}
\end{figure}

The trained NN model based on an ensemble of 100 networks was used to estimate the temperature profiles of all 910 samples in the test data-set. In this study, the ground truth temperature profiles represent a 30 minutes integration time and have a vertical resolution of 30\,m in a height range from 0.7\,km to 6\,km. The retrieved profiles were compared against the ground truth temperature profiles. Fig.~\ref{MBE} (left panel) shows the corresponding mean bias error (MBE) which was calculated as: $\frac{1}{n}\sum_{i}^n (T_{{NN}_{i}} - T_{{truth}_{i}})$. The MBE is always smaller than 0.05\,K for altitude range of 0.7\,km to 4.5\,km and the bias of less than 0.08\,K for the altitude range of 4.5\,km to 6\,km. The bias is much smaller than the average of the summation of the statistical uncertainty of the target profiles ($\frac{1}{N}\sum_{j} \sigma_j$, where N is the number of test profiles) that covers the shaded area. We also calculated the root mean squared error (RMSE) for the test period. The result is shown in Fig.~\ref{MBE} (right panel).

\begin{figure}[htb!]
\centering
\includegraphics[width= 6.5cm, height = 5.0 cm]{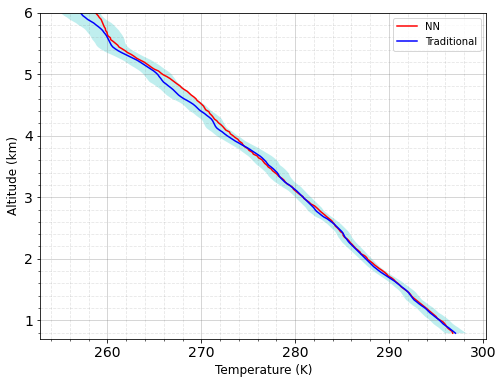}
\includegraphics[width= 6.5cm, height = 5.0 cm]{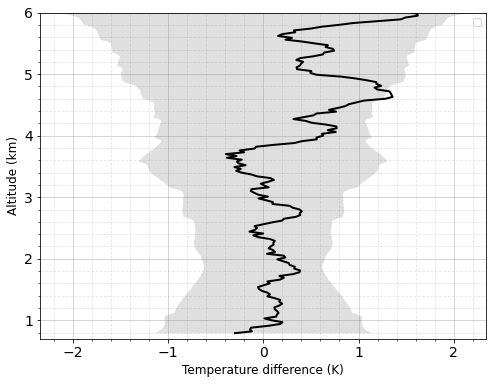}

\caption{The temperature estimation on 23 August 2017 at 22:30 UT. Left panel: The estimated (NN) temperature profile is shown in red, and the target profile is shown in blue. Right panel: The difference between the estimation and the ground truth profiles are plotted (black curve). The gray shaded areas represents the summation of the uncertainties of NN and target profiles.}
\label{T-clear-night}
\end{figure}

An individual NN algorithm is used to retrieve the temperature profiles for both daytime and nighttime measurements, in both clear and cloudy conditions, where the clouds are thin enough for lidar returns to be collected. The NN's estimation as well as the ground truth (traditional) profiles corresponding to 30\,min coadded count measurements from 22:30 UT on 23 August 2017, typical for nighttime clear sky, are shown on the left panel of Fig.~\ref{T-clear-night}. The cyan shaded area shows the summation of uncertainties of the two profiles. The two profiles are inside (within) the shaded area. To see the difference between the two profiles more clearly, ($T_{NN} - T_{traditional}$) is plotted in the right panel of Fig.~\ref{T-clear-night}. The difference between the two profiles is within the summation of the uncertainties of two profiles (shown as the shaded gray area). 

%
%

\begin{figure}[htb!]
\centering
\includegraphics[width= 6.5cm, height = 5.0 cm]{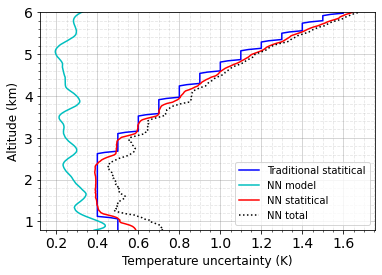}

\caption{The statistical uncertainty of NN's retrieval (red curve) is plotted against the statistical uncertainty of the target profile (blue curve). The model uncertainty is shown in blue, and the total uncertainty is shown with black dotted curve.}
\label{error-clear-night}
\end{figure}

The total uncertainty of NN is also plotted against the uncertainty of the target profile. The total uncertainty on NN (black dotted curve) is the sum of the estimation of the model uncertainty (cyan curve) and the estimated uncertainty of the ground truth (red curve). The model uncertainty does not exceed 0.4\,k at lower altitudes. The estimated uncertainty of the ground truth (red curve) also matches well with the ground truth uncertainty of the ground truth (blue curve). 

\begin{figure}[htb!]
\includegraphics[scale=.30]{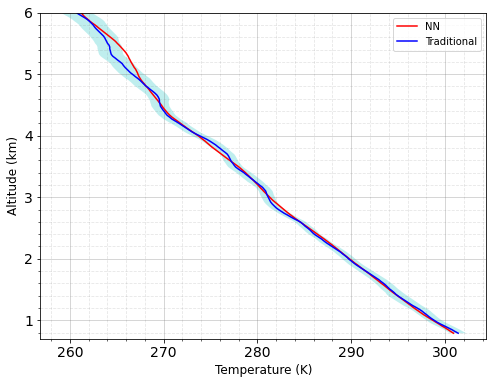}
\includegraphics[scale=.30]{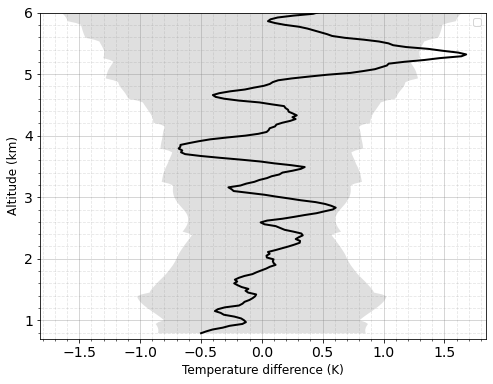}
\includegraphics[scale=.43]{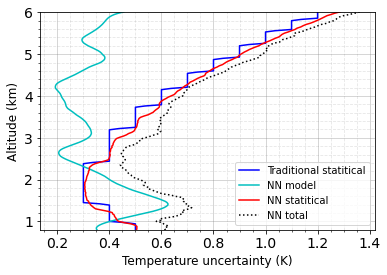}

\caption{The temperature estimation on 6 July 2017 at 16:30 UT. Left panel: The NN estimation (red curve) is plotted against the traditional profile (blue curve). Middle panel: The difference between the estimation and the ground truth profiles are plotted (black curve). Right panel: The statistical uncertainty of NN's retrieval (red curve) is plotted against the statistical uncertainty of the target profile (blue curve). The model uncertainty is shown in blue, and the total uncertainty is shown with black dotted curve.}
\label{clear-day}
\end{figure}

We then considered a typical daytime measurement in clear sky. The traditional and NN temperature profiles of measurements from 16:30 UT on 6 July 2017 are shown in the left panel of Fig.~\ref{clear-day}. The difference between the two profiles is within the summation of their uncertainties. The total uncertainty of the NN is also plotted (right panel of Fig.~\ref{clear-day}). The estimated statistical uncertainty of the NN matches well with the true uncertainty of the traditional profile. The model uncertainty has a low value for most altitudes, however, at about 1.8\,km it shows a higher uncertainty of 
0.6\,K. 


\begin{figure}[htb!]
\centering
\includegraphics[width= 6cm, height = 4.5 cm]{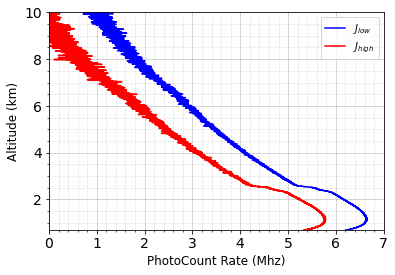}
\caption{measurements from 19:00 UT on 12 Sep 2017; cloudy night.}
\label{cloud-night}
\end{figure}

\begin{figure}[htb!]
\includegraphics[scale=.30]{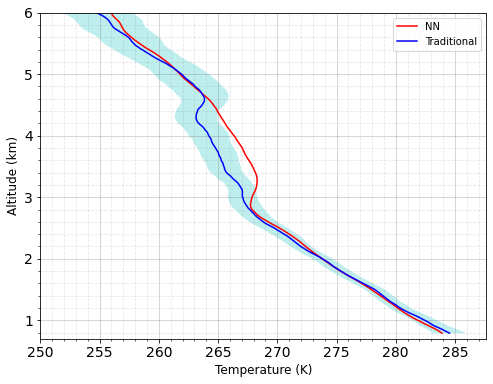}
\includegraphics[scale=.30]{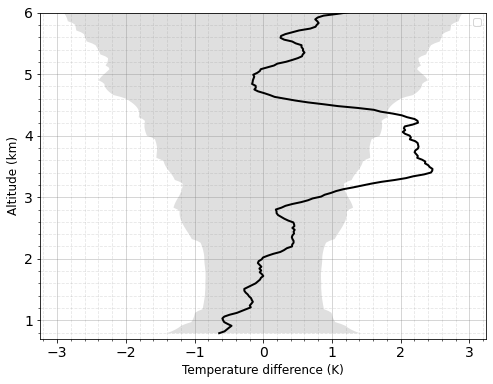}
\includegraphics[scale=.43]{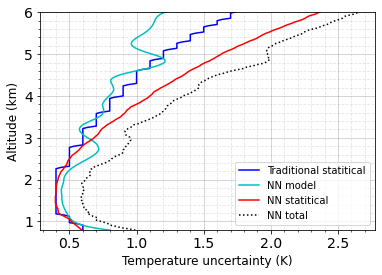}
\caption{The temperature estimation on 6 July 2017 at 19:00 UT. Left plot: The NN estimation (red curve) is plotted against the traditional profile (blue curve). Middle panel: The difference between the estimation and the ground truth profiles are plotted (black curve). Right panel: The statistical uncertainty of NN's retrieval (red curve) is plotted against the statistical uncertainty of the target profile (blue curve). The model uncertainty is shown in blue, and the total uncertainty is shown with the black dotted curve.}
\label{cloud-envelop_night}
\end{figure}

In general, in clear conditions, the agreement between the NN estimations and the traditional profiles are within their uncertainties. However, for cloudy cases, the difference between the two profiles can grow larger. Measurements from 19:00 UT on 12 Sep 2017 indicate that at lower altitude (below 3\,km) a layer of cloud was present (Fig.~\ref{cloud-night}). The NN's estimation as well as the traditional temperature profiles are in good agreement in lower altitudes. At about 3\,km their difference grows larger and at about 4.5\,km the two profiles once again become similar. Left panel in Fig.~\ref{cloud-envelop_night} shows the retrieved NN temperature in red and the traditional profile in blue, also the middle panel, plots the difference between the two profiles. The model and statistical uncertainties of NN profile compared to the uncertainty of NNs in the previous cases (clear night and day) are much larger. However, below 3\,km ( in the altitude range where the agreement between the NN and the traditional profile is good) the uncertainty of NN is smaller than 0.9\,K for most altitudes (Fig.~\ref{cloud-envelop_night} right panel). 

\begin{figure}[htb!]
\includegraphics[scale=.30]{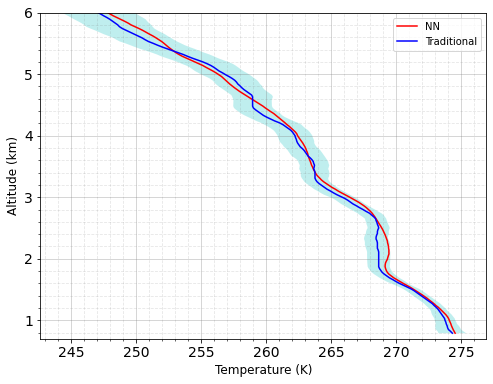}
\includegraphics[scale=.30]{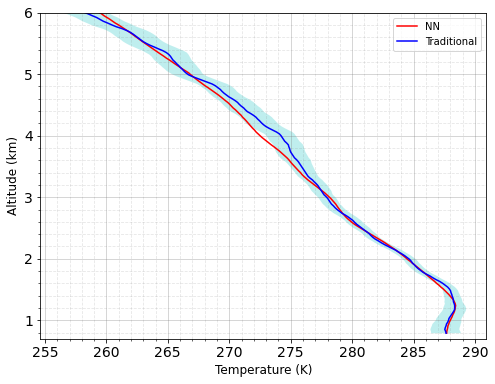}
\includegraphics[scale=.30]{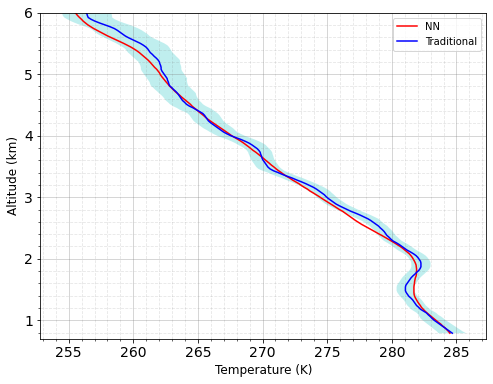}
\caption{Three examples of the NN method successfully retrieving temperature when inversions are present. Left panel: NN and traditional temperature profiles for measurements from 2:30 UT on 5 December 2017. Middle panel: NN and traditional temperature profiles for measurements from 11:30 UT on 26 October 2017. Right panel: NN and traditional temperature profiles for measurements from 19:00 UT on 3 November 2017.}
\label{other_examples}
\end{figure}

To further illustrate our method we consider 3 interesting cases with moderate to large temperature inversions. Fig.~\ref{other_examples} shows more examples of estimated NN temperatures plotted against their corresponding traditional temperature profiles. The left panel shows the result for measurements from 2:30 UT on 5 December 2017. The traditional temperature profile shows significant variability, and the NN estimation can successfully capture these temperature changes. The middle panel represents temperature profiles from 11:30 UT on 26 October 2017. Similar to the previous example, the NN estimation is in good agreement with the traditional profile. The right panel shows results for measurements from 19:00 UT on 3 November 2017. In this night, a thin layer of clouds was presented at higher altitudes (above 4.5\,km). 

\section{Summary and Conclusions}
Our study shows proof-of-concept that NNs are capable of estimating the ground truth temperature profiles. We have shown that a single model can be trained to estimate both daytime and nighttime temperature profiles in various atmospheric conditions. We trained our model using 3600 measurement and their corresponding temperature profiles. We evaluated our model using 910 unseen measurements. The corresponding MBE exhibited a bias of less than 0.05\,K in the range of estimation. We showed that the trained model is capable of estimating temperature profiles for different atmospheric conditions (clear and cloudy). However, for cloudy conditions, at some altitudes, the difference between the NN model and the traditional could grew larger than the clean conditioned estimations. Details of retrieved profiles for a few case studies were also provided.

Another major focus of the current study was to provide a reliable uncertainty budget for retrieved profiles. One of the ongoing issues with deep learning models is that they are overconfident, meaning that in the models' estimations little or no uncertainty is considered. However, in reality, the optimized weights are uncertain and their uncertainty can highly affect the estimation of the models. Here, we approached the problem by training an ensemble of networks each of which trained to estimate a MAP estimation. The distribution of the solutions is a good estimate of the true distribution. Thus, we could use the temperature profiles and produce reliable model uncertainties. In general, for clear atmospheric conditions, the NN's retrieval is within the uncertainty of the traditional method. The estimated statistical uncertainty of the NN matches well with the ground truth uncertainty, and the model uncertainty is small. It is important to mention, that unlike the inherited uncertainty from the measurements, the model uncertainty can be minimized by training the model on larger data-sets. For example, as the cloudy conditions are more difficult cases to learn by the algorithm, the model uncertainty shows larger uncertainties. Thus training the model with more data can result in better estimations of the model uncertainty. Hence, one future step is to train the model with a larger data-set to investigate if the temperature profiles from the cloudy measurements can become closer to the ground truth. Moreover, in this study, we trained a general model for day-time, night-time, and cloudy conditions. Granting a larger data-set, it is interesting to build three condition-dependent models and compare the outputs with the output of the general model. Another emerging field of research is multitask learning. Implementing this method, the input of the network contains measurements corresponding to different conditions, and the first few layers of the network are shared between all types of measurements, but the last few layers are trained specifically for each of the mentioned conditions \cite{caruana1997multitask}. The comparison of the accuracy of the estimated profiles from training the NNs based on the general, condition-dependent and multitask models will help us to determine the best model for our application. 

Typically, NNs are data-driven and require minimal data processing. For example, there is no need to use any additional data preprocessing such as correcting the deadtime effect, calculating the overlap function and etc. However, in this study, we have used the retrieved RALMO temperature profile as the ground truth. Although, the NN algorithm is free of data processing steps, to provide the ground truth extensive data processing is required. Thus, using RALMO temperature profiles as the ground truth is one of the limitations of the present study. Hence, one of the major future steps is to use radiosonde temperature profiles as the ground truth such that the estimated profiles become independent of RALMO historical temperature profiles.

We are also planning to explore the possibility of training a Bayesian NN from other rotational Raman lidars. This way we can have a single trained model capable of estimating temperature profiles from different lidars. One advantage of developing such an algorithm is the ease of comparing temperature profiles from different locations. After the initial training process, implementing the model on new unseen data is quite fast, estimating temperature profiles, even on an old laptop, takes only a few seconds. Considering the reliable result we achieved for the temperature profiles, we are interested to use the NN algorithm to estimate the water vapor profiles from RALMO. 


\bibliographystyle{unsrt}
{\footnotesize
\bibliography{PINN.bib}}

\begin{thebibliography}{10}

\bibitem{santer1996search}
Benjamin~D Santer, KE~Taylor, TML Wigley, TC~Johns, PD~Jones, DJ~Karoly, JFB
  Mitchell, AH~Oort, JE~Penner, V~Ramaswamy, et~al.
\newblock A search for human influences on the thermal structure of the
  atmosphere.
\newblock {\em Nature}, 382(6586):39--46, 1996.

\bibitem{tett1996human}
Simon~FB Tett, John~FB Mitchell, David~E Parker, and Myles~R Allen.
\newblock Human influence on the atmospheric vertical temperature structure:
  Detection and observations.
\newblock {\em Science}, 274(5290):1170--1173, 1996.

\bibitem{thorne2011tropospheric}
Peter~W Thorne, John~R Lanzante, Thomas~C Peterson, Dian~J Seidel, and Keith~P
  Shine.
\newblock Tropospheric temperature trends: History of an ongoing controversy.
\newblock {\em Wiley Interdisciplinary Reviews: Climate Change}, 2(1):66--88,
  2011.

\bibitem{philipona2018radiosondes}
Rolf Philipona, Carl Mears, Masatomo Fujiwara, Pierre Jeannet, Peter Thorne,
  Greg Bodeker, Leopold Haimberger, Maxime Hervo, Christoph Popp, Gonzague
  Romanens, et~al.
\newblock Radiosondes show that after decades of cooling, the lower
  stratosphere is now warming.
\newblock {\em Journal of Geophysical Research: Atmospheres}, 123(22):12--509,
  2018.

\bibitem{cooney1972measurement}
John Cooney.
\newblock Measurement of atmospheric temperature profiles by raman backscatter.
\newblock {\em Journal of Applied Meteorology and Climatology}, 11(1):108--112,
  1972.

\bibitem{behrendt2005temperature}
Andreas Behrendt.
\newblock Temperature measurements with lidar.
\newblock In {\em Lidar}, pages 273--305. Springer, 2005.

\bibitem{mahagammulla2019retrieval}
Shayamila Mahagammulla~Gamage, Robert~J Sica, Giovanni Martucci, and Alexander
  Haefele.
\newblock Retrieval of temperature from a multiple channel pure rotational
  raman backscatter lidar using an optimal estimation method.
\newblock {\em Atmospheric Measurement Techniques}, 12(11):5801--5816, 2019.

\bibitem{pierson2017deep}
Harry~A Pierson and Michael~S Gashler.
\newblock Deep learning in robotics: a review of recent research.
\newblock {\em Advanced Robotics}, 31(16):821--835, 2017.

\bibitem{duc20203d}
Nguyen~Thanh Duc, Seungjun Ryu, Muhammad Naveed~Iqbal Qureshi, Min Choi, Kun~Ho
  Lee, and Boreom Lee.
\newblock 3d-deep learning based automatic diagnosis of alzheimer’s disease
  with joint mmse prediction using resting-state fmri.
\newblock {\em Neuroinformatics}, 18(1):71--86, 2020.

\bibitem{yazdi2020sefee}
Amirhessam Yazdi, Xing Lin, Lei Yang, and Feng Yan.
\newblock Sefee: lightweight storage error forecasting in large-scale
  enterprise storage systems.
\newblock In {\em 2020 SC20: International Conference for High Performance
  Computing, Networking, Storage and Analysis (SC)}, pages 894--907. IEEE
  Computer Society, 2020.

\bibitem{sun2020surrogate}
Luning Sun, Han Gao, Shaowu Pan, and Jian-Xun Wang.
\newblock Surrogate modeling for fluid flows based on physics-constrained deep
  learning without simulation data.
\newblock {\em Computer Methods in Applied Mechanics and Engineering},
  361:112732, 2020.

\bibitem{muller2003ozone}
Martin~D M{\"u}ller, Anton~K Kaifel, Mark Weber, Silvia Tellmann, John~P
  Burrows, and Diego Loyola.
\newblock Ozone profile retrieval from global ozone monitoring experiment
  (gome) data using a neural network approach (neural network ozone retrieval
  system (nnorsy)).
\newblock {\em Journal of Geophysical Research: Atmospheres}, 108(D16), 2003.

\bibitem{blackwell2005neural}
William~J Blackwell.
\newblock A neural-network technique for the retrieval of atmospheric
  temperature and moisture profiles from high spectral resolution sounding
  data.
\newblock {\em IEEE Transactions on Geoscience and Remote Sensing},
  43(11):2535--2546, 2005.

\bibitem{del2005neural}
F~Del~Frate, M~Iapaolo, S~Casadio, Sophie Godin-Beekmann, and Monique
  Petitdidier.
\newblock Neural networks for the dimensionality reduction of gome measurement
  vector in the estimation of ozone profiles.
\newblock {\em Journal of Quantitative Spectroscopy and Radiative Transfer},
  92(3):275--291, 2005.

\bibitem{milstein2016neural}
Adam~B Milstein and William~J Blackwell.
\newblock Neural network temperature and moisture retrieval algorithm
  validation for airs/amsu and cris/atms.
\newblock {\em Journal of Geophysical Research: Atmospheres},
  121(4):1414--1430, 2016.

\bibitem{cai2020temperature}
Xi~Cai, Yansong Bao, George~P Petropoulos, Feng Lu, Qifeng Lu, Liuhua Zhu, and
  Ying Wu.
\newblock Temperature and humidity profile retrieval from fy4-giirs
  hyperspectral data using artificial neural networks.
\newblock {\em Remote Sensing}, 12(11):1872, 2020.

\bibitem{tiwari2010uncertainty}
Mukesh~Kumar Tiwari and Chandranath Chatterjee.
\newblock Uncertainty assessment and ensemble flood forecasting using bootstrap
  based artificial neural networks (banns).
\newblock {\em Journal of Hydrology}, 382(1-4):20--33, 2010.

\bibitem{blundell2015weight}
Charles Blundell, Julien Cornebise, Koray Kavukcuoglu, and Daan Wierstra.
\newblock Weight uncertainty in neural network.
\newblock In {\em International Conference on Machine Learning}, pages
  1613--1622. PMLR, 2015.

\bibitem{gal2016dropout}
Yarin Gal and Zoubin Ghahramani.
\newblock Dropout as a bayesian approximation: Representing model uncertainty
  in deep learning.
\newblock In {\em international conference on machine learning}, pages
  1050--1059. PMLR, 2016.

\bibitem{lakshminarayanan2016simple}
Balaji Lakshminarayanan, Alexander Pritzel, Charles Blundell, and London
  DeepMind.
\newblock Simple and scalable predictive uncertainty estimation using deep
  ensembles.
\newblock {\em stat}, 1050:5, 2016.

\bibitem{pearce2020uncertainty}
Tim Pearce, Felix Leibfried, and Alexandra Brintrup.
\newblock Uncertainty in neural networks: Approximately bayesian ensembling.
\newblock In {\em International conference on artificial intelligence and
  statistics}, pages 234--244. PMLR, 2020.

\bibitem{leblanc2016proposed}
Thierry Leblanc, Robert~J Sica, Joanna~AE van Gijsel, Alexander Haefele,
  Guillaume Payen, and Gianluigi Liberti.
\newblock Proposed standardized definitions for vertical resolution and
  uncertainty in the ndacc lidar ozone and temperature algorithms--part 3:
  Temperature uncertainty budget.
\newblock {\em Atmospheric Measurement Techniques}, 9(8):4079--4101, 2016.

\bibitem{zhang2014slope}
Yunpeng Zhang, Fan Yi, Wei Kong, and Yang Yi.
\newblock Slope characterization in combining analog and photon count data from
  atmospheric lidar measurements.
\newblock {\em Applied optics}, 53(31):7312--7320, 2014.

\bibitem{pettifer1975signal}
REW Pettifer.
\newblock Signal induced noise in lidar experiments.
\newblock {\em Journal of Atmospheric and Terrestrial Physics}, 37(4):669--673,
  1975.

\bibitem{acharya2004signal}
YB~Acharya, Som Sharma, and H~Chandra.
\newblock Signal induced noise in pmt detection of lidar signals.
\newblock {\em Measurement}, 35(3):269--276, 2004.

\bibitem{wandinger2002experimental}
Ulla Wandinger and Albert Ansmann.
\newblock Experimental determination of the lidar overlap profile with raman
  lidar.
\newblock {\em Applied Optics}, 41(3):511--514, 2002.

\bibitem{hervo2016empirical}
Maxime Hervo, Yann Poltera, and Alexander Haefele.
\newblock An empirical method to correct for temperature-dependent variations
  in the overlap function of chm15k ceilometers.
\newblock {\em Atmospheric Measurement Techniques}, 9(7):2947--2959, 2016.

\bibitem{dinoev2013raman}
T~Dinoev, V~Simeonov, Y~Arshinov, S~Bobrovnikov, Pablo Ristori, B~Calpini,
  M~Parlange, and H~Bergh.
\newblock Raman lidar for meteorological observations, ralmo--part 1:
  Instrument description.
\newblock {\em Atmospheric Measurement Techniques}, 6(5):1329--1346, 2013.

\bibitem{martucci2021validation}
Giovanni Martucci, Francisco Navas-Guzm{\'a}n, Ludovic Renaud, Gonzague
  Romanens, S~Mahagammulla Gamage, Maxime Hervo, Pierre Jeannet, and Alexander
  Haefele.
\newblock Validation of pure rotational raman temperature data from the raman
  lidar for meteorological observations (ralmo) at payerne.
\newblock {\em Atmospheric Measurement Techniques}, 14(2):1333--1353, 2021.

\bibitem{zhigljavsky2007stochastic}
Anatoly Zhigljavsky and Antanasz Zilinskas.
\newblock {\em Stochastic global optimization}, volume~9.
\newblock Springer Science \& Business Media, 2007.

\bibitem{torn1999stochastic}
Aimo T{\"o}rn, Montaz~M Ali, and Sami Viitanen.
\newblock Stochastic global optimization: Problem classes and solution
  techniques.
\newblock {\em Journal of Global Optimization}, 14(4):437--447, 1999.

\bibitem{kingma2014adam}
Diederik~P Kingma and Jimmy Ba.
\newblock Adam: A method for stochastic optimization.
\newblock {\em International Conference on Learning Representations (ICLR)},
  2015.

\bibitem{bishop1995training}
Chris~M Bishop.
\newblock Training with noise is equivalent to tikhonov regularization.
\newblock {\em Neural computation}, 7(1):108--116, 1995.

\bibitem{theodoridis2015machine}
Sergios Theodoridis.
\newblock {\em Machine learning: a Bayesian and optimization perspective}.
\newblock Academic press, 2015.

\bibitem{quinonero2005evaluating}
Joaquin Quinonero-Candela, Carl~Edward Rasmussen, Fabian Sinz, Olivier
  Bousquet, and Bernhard Sch{\"o}lkopf.
\newblock Evaluating predictive uncertainty challenge.
\newblock In {\em Machine Learning Challenges Workshop}, pages 1--27. Springer,
  2005.

\bibitem{welling2011bayesian}
Max Welling and Yee~W Teh.
\newblock Bayesian learning via stochastic gradient langevin dynamics.
\newblock In {\em Proceedings of the 28th international conference on machine
  learning (ICML-11)}, pages 681--688. Citeseer, 2011.

\bibitem{korattikara2015bayesian}
Anoop Korattikara, Vivek Rathod, Kevin Murphy, and Max Welling.
\newblock Bayesian dark knowledge.
\newblock {\em arXiv preprint arXiv:1506.04416}, 2015.

\bibitem{hmcspringenberg2016bayesian}
Jost~Tobias Springenberg, Aaron Klein, Stefan Falkner, and Frank Hutter.
\newblock Bayesian optimization with robust bayesian neural networks.
\newblock In {\em Proceedings of the 30th International Conference on Neural
  Information Processing Systems}, pages 4141--4149, 2016.

\bibitem{bardsley2014randomize}
Johnathan~M Bardsley, Antti Solonen, Heikki Haario, and Marko Laine.
\newblock Randomize-then-optimize: A method for sampling from posterior
  distributions in nonlinear inverse problems.
\newblock {\em SIAM Journal on Scientific Computing}, 36(4):A1895--A1910, 2014.

\bibitem{liu2006estimating}
Zhaoyan Liu, William Hunt, Mark Vaughan, Chris Hostetler, Matthew McGill,
  Kathleen Powell, David Winker, and Yongxiang Hu.
\newblock Estimating random errors due to shot noise in backscatter lidar
  observations.
\newblock {\em Applied optics}, 45(18):4437--4447, 2006.

\bibitem{sengupta2020ensembling}
Ushnish Sengupta, Matt Amos, J~Scott Hosking, Carl~Edward Rasmussen, Matthew
  Juniper, and Paul~J Young.
\newblock Ensembling geophysical models with bayesian neural networks.
\newblock {\em arXiv preprint arXiv:2010.03561}, 2020.

\bibitem{caruana1997multitask}
Rich Caruana.
\newblock Multitask learning.
\newblock {\em Machine learning}, 28(1):41--75, 1997.

\end{thebibliography}

\end{document}